# Improving Data Quality in Intelligent Transportation Systems

V.M. Megler, Kristin Tufte, and David Maier, *Senior Member, IEEE*

*Abstract*—**Intelligent Transportation Systems (ITS) use data and information technology to improve the operation of our transportation network. ITS contributes to sustainable development by using technology to make the transportation system more efficient; improving our environment by reducing emissions, reducing the need for new construction and improving our daily lives through reduced congestion. A key component of ITS is traveler information. The Oregon Department of Transportation (ODOT) recently implemented a new traveler information system on selected freeways to provide drivers with travel time estimates that allow them to make more informed decisions about routing to their destinations. The ODOT project aims to improve traffic flow and promote efficient traffic movement, which can reduce emissions rates and improve air quality.**

**The new ODOT system is based on travel data collected from a recently-increased set of sensors installed on its freeways. Our current project investigates novel data cleaning methodologies and the integration of those methodologies into the prediction of travel times. We use machine learning techniques on our archive to identify suspect data, and calculate revised travel times excluding this suspect data. We compare the resulting travel time predictions to ground-truth data, and to predictions based on simple, rule-based data cleaning. We report on the results of our study using qualitative and quantitative methods.**

## I. INTRODUCTION

INTELLIGENT Transportation Systems (ITS) aim to use data and technology to improve the safety, efficiency and reliability of our transportation system. Given recent environmental predictions of global climate change and the impact of transportation on greenhouse gas production, ITS as a mechanism for reducing emissions via technology are growing more and more important. Consider that transportation accounts for 27% of greenhouse gasses produced in the United States [1] and that — further — in 2012 in the United States, commuters spent 5.5 billion hours in congested traffic, wasting 2.9 billion gallons of fuel and producing 56 billion lbs. of CO2 [2]. Transportation has a significant impact on the sustainability of our environment.

Typically, ITS include traveler information in the form of travel times posted on Variable Message Signs (VMS) on major roads to help drivers make more informed decisions about their travel route choices. Traveler information, such as travel times, is dependent both on the quality of the prediction algorithms and on the quality of the data cleaning applied to the data used by the prediction algorithms. Sensor data is known to be dirty and must be cleaned prior to use. Existing cleaning approaches are often rule-based and relatively static. We were curious whether machine learning techniques could be applied to further improve data cleaning, and whether the results of applying such methods would meaningfully alter the results of analyses performed on the data.

We use data from the Portal data archive [3] — a 3TB data of transportation-related data containing traffic sensor data for Portland, Oregon freeways — along with ground truth to test machine-learning based data cleaning. We apply clustering techniques to the raw Portal data to identify the main traffic characteristics of different traffic patterns, such as congestion or light traffic (here, called "regimes"). We then identify outliers for each traffic cluster, and assert that these outliers are "bad data." Lastly we apply the results of this flexible cleaning approach to calculating travel times, and compare them to a ground truth travel time dataset.

## II. DATA CLEANING FOR TRAFFIC SENSOR DATA

Sensor data cleaning is an important issue, as data products are only as good as the input data they are based on. We observe that different data products require different levels of data quality and different analyses require different cleaning processes. For example, it is clear that a different level of cleaning and quality is required for producing a speed map versus research into bottleneck identification.

Current data cleaning approaches tend to assume that "bad data" should be removed from the raw data and a "clean" data set created. Analyses are then performed using the "clean data." It further assumes that "bad data" can be recognized as such. However, the definition of what is bad data is often dependent on the actual analysis to be performed. We contrast traveler information systems, performance measure reports and sensor failure detection. For real-time traveler information (e.g., a speed map, variable advisory speed signs or travel time signs), data can only be evaluated in the context of data collected prior to "now." A high priority may be put on having information at most times as opposed to removing all suspect information, and long gaps in cleaned data are not acceptable. Further, the data cleaning process must be efficient enough to occur in real-time in an Advanced Traffic Management System (ATMS).

In contrast, data for performance measure reports can be processed offline and data can be compared to data collected both before and after the data item to be cleaned. Here, clean data may have a higher priority than having continuous data. In addition, traveler information and performance reports

This paper was first submitted on August 28, 2015. This work was supported in part by Intel Science and Technology Center for Big Data and the Maseeh Professorship in Emerging Technologies.

This research was performed while V.M. Megler (e-mail: vmegler@gmail.com) was at Portland State University. Kristin Tufte (e-mail: tufte@pdx.edu) and David Maier (e-mail: maier@pdx.edu) are with Portland State University, Portland, OR 97201 USA.

often want to include data from days with incidents in their analysis, while researchers and planners often want to exclude incidents as they focus on "the common case." Lastly, a person trying to understand sensor failures will want outliers included in their results.

In practice, each analyst tends to perform multiple analyses, selecting subsets of data, performing calculations, checking the results, excluding suspect sections, and re-running. In essence, data cleaning is really an iterative process, with the specific use case driving what is defined as "clean data." In our work with other data archives, we find the same practice of iterative data cleaning [4], [5]. We therefore challenge the concept of a one-time data cleaning step, and propose instead a more flexible view of repeated data cleaning as a central part of the analytic process.

At the high level, traffic can be broken into (at least) three "regimes" with different characteristics. First, free flow-low volume, as generally occurs at night. Here, traffic is flowing freely at posted speeds and there are few cars. Secondly, free-flow-medium volume: traffic flows freely but with a higher volume of cars, such as may occur just before or after congestion or in the middle of the day. Finally, congestion, where traffic volume is high and speeds are low. (Further segmentation is also possible.) While traffic regimes are often at least somewhat predictable (congestion during peak hours, for example), they may often occur at other times due to specific events (a traffic incident, or a popular concert causing congestion at an unusual time). In addition, freeway congestion for, e.g., the afternoon peak traffic (assumed to be 4 p.m. to 6 p.m.) may begin at a different time (say, 3:30 p.m.) from the assumed one. Incidents can also cause a shift from the current or expected traffic regime (say, light traffic) to another (congested). Since the different regimes have different characteristics, data may need different cleaning rules for each regime. For example: a traffic speed reading of 0 with high occupancy (the percentage of time that there is a vehicle over the detector) may be a valid reading during high congestion, but is unlikely to be so during free-flowing traffic. We would like a data cleaning approach that can adapt to these nuances.

## III. RELATED WORK

We describe three kinds of related work: data quality work in the domain of traffic data; use of machine learning for data quality; and clustering and anomaly detection.

### A. Data Quality in Traffic Data

Original research in data quality for ITS data was performed by Turner, et al. [6]. They proposed a set of tests including threshold tests such as 20-second volume > 17. The threshold tests applied to individual 20-second data records and indicated if those individual records were good or bad. The tests were based on unlikely readings; for example a speed > 100 or a 20-second volume > 17 is much more likely to be a detector error than to be a valid reading. In a similar manner, Turochy and Smith [7] proposed a set of five tests declaring a data record "bad" if it failed one or more of those tests.

There has not been much recent work aimed at improving on these methods, as Hamad and Quiroga note in their report on data quality for TransGuide [8]. Much transportation data-quality checking is still based on rule-based tests such as those initially proposed by Turner et al. and Turochy and Smith. In fact, Hamad and Quiroga generally follow those threshold (and other) rule-based tests.

A different method of data quality filtering was used by the Seattle Federal project [9]. This project combined several types of data quality filtering. First, for the 20-second granularity data, Gaussian Mixture Models are used to find poorly-performing detectors, based on a paper by Corey, et al. [10]. Once a poorly-performing detector is identified, the entire month's data for that detector is removed. For 5-minute aggregations, the Seattle Federal project uses a set of rules including data gaps and the typical high speed, low volume rules to assign a data quality flag. This assignment is followed by human review, as necessary; and finally, data is flagged as "good" or "bad." As with the 20-second data, for data identified as bad, all data for that month is removed. This month-level removal is acceptable for performance metric calculation. In contrast, our work focuses on real-time travel time calculations, so we cannot wait for human review. We also cannot remove the current month of data when bad data is detected, as we must continue to post travel times. We seek a finer-grained detection approach.

Other data quality research includes quality assessment (e.g. Huber, et al. [11]) and data imputation (e.g. Henrickson, et al. [12] and Jie, et al. [13]). These techniques are complementary to our approach and applied after our data quality filtering.

### B. Anomaly Detection

Clustering is one of the most popular unsupervised learning approaches. As defined by Jain [14], the goal of data clustering is to discover the natural grouping(s) of a set of patterns, points, or objects. He provides an operational definition of clustering: Given a representation of $n$ objects, find $K$ groups based on a measure of similarity such that the similarities between objects in the same group are high while the similarities between objects in different groups are low. As Jain points out, a cluster is a subjective entity whose significance and interpretation requires domain knowledge [14]. In our case, we wished to identify the natural grouping, or partitions, of traffic speed, volume and occupancy into different traffic regimes to help recognize anomalous readings.

Chandola et al. define three categories of anomalies: point anomalies, where an individual data instance is anomalous with respect to the rest of the data; contextual anomalies, where a data instance is anomalous in a specific context (but not otherwise); and collective anomalies, where the co-occurrence of a set of data instances is the anomaly [15]. Ideally, we would like to identify all three kinds of anomalies in our traffic data.

Anomaly detection methods based on clustering can be roughly broken into two categories: classification and outlier detection [16]. Classification methods focus on identifying a cluster of similar but anomalous objects. They assume that a significant set of anomalies are more similar to each other than to other, non-anomalous objects; that is, the anomalies will form a cluster of their own. Outlier detection assumes that objects that are further than some distance $d$ from any cluster center are anomalous. Classification and outlier detection can be combined (as in Münz et al. [16]). If the two methods are applied simultaneously, an object is treated as an anomaly if it is closer to an anomalous cluster centroid than to a normal one, or if its distance to a normal cluster centroid is larger than some chosen distance.

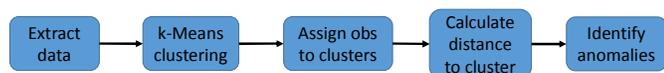

Fig. 1. Traffic-observation data-cleaning pipeline. Anomalies are detected based on clustering and on Mahalanobis distance to the closest cluster center.

Anomaly detection has been used extensively in identifying network intrusions (e.g., [15]–[19]). Patcha and Park [18] provide an overview of a number of methods that have been used, including Bayesian networks, hidden Markov models, principal components analysis (using Mahalanobis distances, amongst others), clustering and outlier detection. Eskin [17] compares three clustering algorithms for detecting outliers: cluster-based estimation; K-nearest neighbor, and one-class Support Vector Machines. Münz et al. use K-means clustering to identify anomalies in computer network traffic [16]; their approach assumes that anomalous data forms a specific cluster, clearly separate from normal network-traffic data. They follow this technique with outlier detection, as do we. However, most other researchers focus either on identifying an anomalous cluster, separate from normal data, or on identifying individual outliers.

Zhang et al. surveys anomaly detection in wireless sensor networks [20]. One of the challenges they draw attention to is how to distinguish between individual outliers (errors) and events (in our domain, for example, a sensor malfunctioning but continuing to send incorrect data). In the domain of anomaly detection for networks of sensors of vehicular traffic, Shekhar et al. apply anomaly detection techniques to a similar network of traffic sensors in Minneapolis-St. Paul; however, they focus on identifying the location of sensor stations whose measurements are inconsistent with those of their topologically-connected neighboring stations [21].

## IV. METHOD

Given that domain experts think of traffic flow in terms of "regimes" with similar characteristics, we felt that clustering methods would be a fruitful machine-learning technique to explore. We postulated that a data-cleaning pipeline as shown in Figure 1 could be effective in applying clustering techniques to traffic-data quality. This section describes the individual steps of that pipeline.

### A. Data Extraction

Our original plan was to apply supervised learning techniques, using existing raw and cleaned data to create two labeled sample sets: "good" (kept after cleaning) and "bad" (discarded during cleaning). However, discussion with domain experts established that while they often have confidence in some specific data values or combinations as being "bad data," for other values they have less confidence. They were curious as to what data values would be identified using machine-learning techniques and how those compared to ones they would identify using traditional techniques. We therefore used unsupervised learning methods, that is, we look for structure in the unlabeled data and use that structure to identify anomalies.

In line with machine-learning conventions we began by identifying a training set [22]. For our training set we used data for the month of May, 2015. May is often used as a "representative month" in Portal traffic analyses, as there are no school holidays and it generally has few adverse weather effects. The selected features were the three variables reported by the sensors: volume (number of vehicles passing the sensor), speed and occupancy (a measure of vehicle density).

We chose five highway segments of interest; these highway segments are experientially known to have different traffic patterns in terms of traffic speeds and volumes, varying from highly-congested (I-5 (South of downtown)) to relatively free-flowing (I-205). Each segment has a north-bound (NB) and south-bound (SB) component, which are treated separately (as they are by the traffic analysts) for a total of ten segments. The number and placement of traffic sensors varies across the segments. In essence, the choice of freeway segments provides a first level of clustering, on a geographic basis. Data counts per freeway segment vary from 2.6M to 4.5M observations for May 2015, depending primarily on the number of detectors.

We extracted detector observations for the chosen traffic segments and month from Portal [23]. The source archive is a PostgreSQL database. We uploaded the data into an Amazon Web Services cluster and processed it using Spark 1.4.1 [24] in Python, with Numpy and Scipy libraries.

### B. Clustering

We chose K-means clustering, the most widely-used algorithm for clustering [14], using the Spark MLLIB Python implementation.

We clustered each freeway segment in two different ways: firstly, using all data for the month. Secondly, we partitioned each segment temporally into groups of similar days-of-the-week (Monday/Friday, Tuesday/Wednesday/Thursday, and Saturday/Sunday). We clustered the data using values of $K$ from 1 to 10, calculating the average squared error for each cluster. We plotted the resulting average square error for each traffic segment and temporal combination (shown in Figure 2), and used the knee-of-the-curve heuristic combined with domain knowledge to select the number of clusters ($K$=3) to be used for the next step. (Note that while we seek to decrease the average squared error, increasing the number of clusters to match the number of data points will always give the lowest squared error; however, that solution has limited utility.)

For the selected number of clusters, we plotted the cluster centers in a series of spider charts and reviewed them with a domain expert.

### C. Assign to Clusters and Calculate Distance

We reprocessed the source data, assigning each observation to the closest cluster and calculating the distance of the observation from the cluster center to which it was assigned. We used the Mahalanobis distance function; this alternative to Euclidean distance is computed based on the sum of squares of the standardized principal component scores. It inherently normalizes the features to account for different units and scales in the observed data, and compensates for statistical correlation between different features. Our domain experts noted that there was some relationship between traffic volumes, occupancy and speeds, although the relationship is not a linear one. Since we had a small number of features and were using an elastic cluster for our processing, the additional computational overhead of calculating and inverting the covariance matrix for our data (an input to the Mahalanobis distance function) [14] was not a significant constraint. Mahalanobis distance is particularly well suited to multivariate data. *Multivariate outliers* are cases that have an unusual combination of values for a number of variables. The value for any of the individual variables may not be an outlier for that variable, however it may be a value that occurs only rarely in combination with the other variable values.

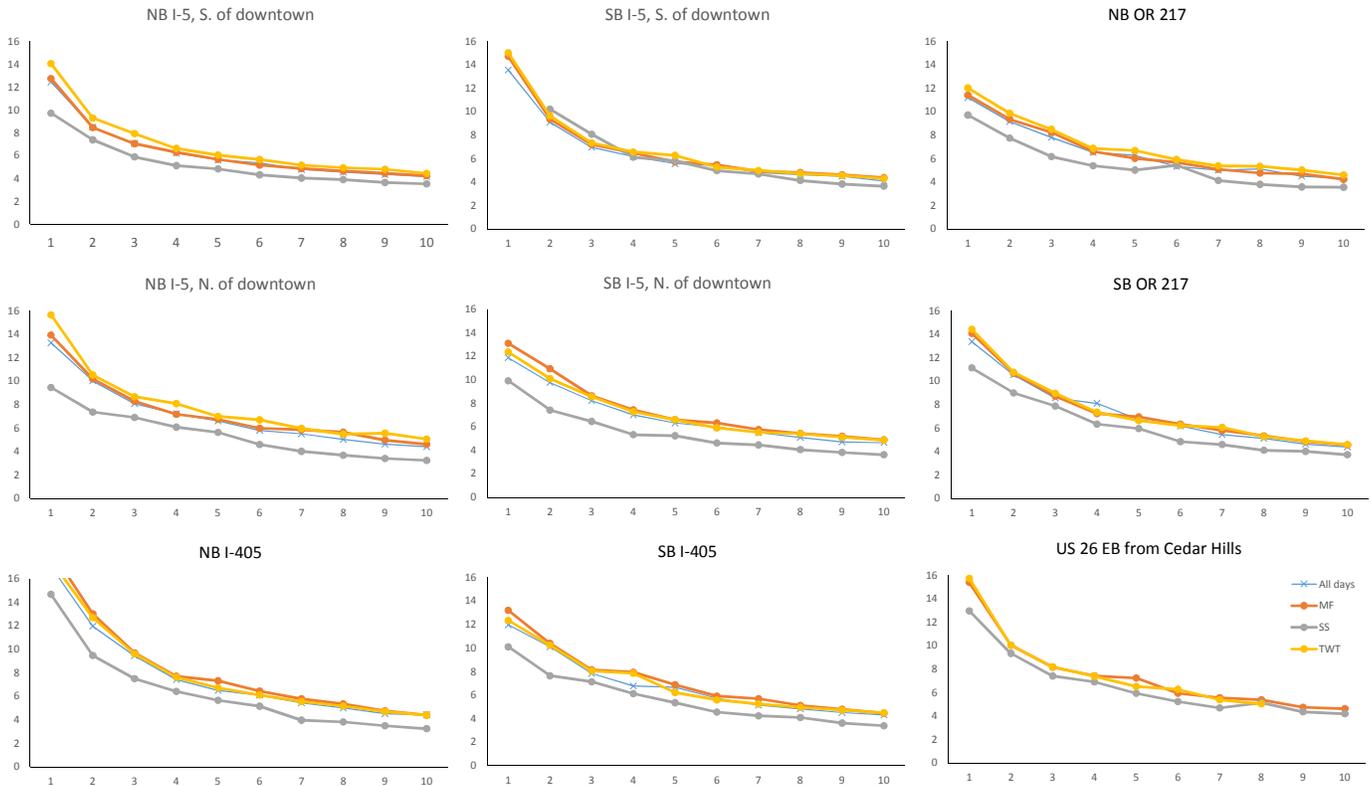

Fig. 2. Average squared error for 1 to 10 clusters for several freeway sections of interest. The freeway segments chosen have very different traffic characteristics. Each freeway segment is plotted separately for Northbound and Southbound traffic, as the traffic patterns alter based on commute directions. Each chart shows four temporal groupings: all days of the week, Monday/Friday, Tuesday/Wednesday/Thursday, and Saturday/Sunday.

### D. Anomaly Detection

We wish to identify two kinds of anomalies: cases where there are enough instances of an anomaly for them to form their own cluster; and outliers, where some observations are unlikely to be valid but are each dissimilar from the others. For the former, we use review of the cluster centers using spider plots such as those shown in Figures 3 and 4 (discussed further below). For outliers that do not themselves form a cluster, we identify all data with a Mahalanobis distance $d > 2.5$ as being anomalies (with $d = 1$ being an analog of 1 standard deviation for each variable dimension).

### E. Validation

We performed several kinds of validation. We used machine-learning tools such as learning charts (see Figure 4), and reviewed statistics of raw and processed data. In addition, given the domain experts' familiarity with their data and their existing analysis tools and views, we relied heavily on their manual review; this approach fit with their existing method of performing some data analytics, then reviewing the results for plausibility. We created plots of raw and corrected data for various time-slices and freeway segments expected to have very different characteristics. Where any results raised questions (such as the anomalous clusters), we researched the underlying data for the cause(s) of that anomaly.

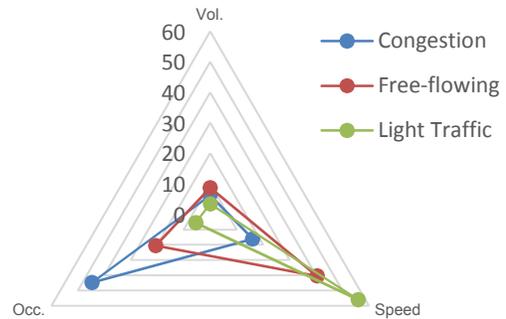

Fig. 3. Traffic Regimes by Day. Each spider plot shows the k=3 clusters for the I-5 NB (South) freeway segment, for a subset of "similar traffic days" (Tuesday/Wednesday/Thursday). The three variables used in the clustering, volume, speed and occupancy, are shown.

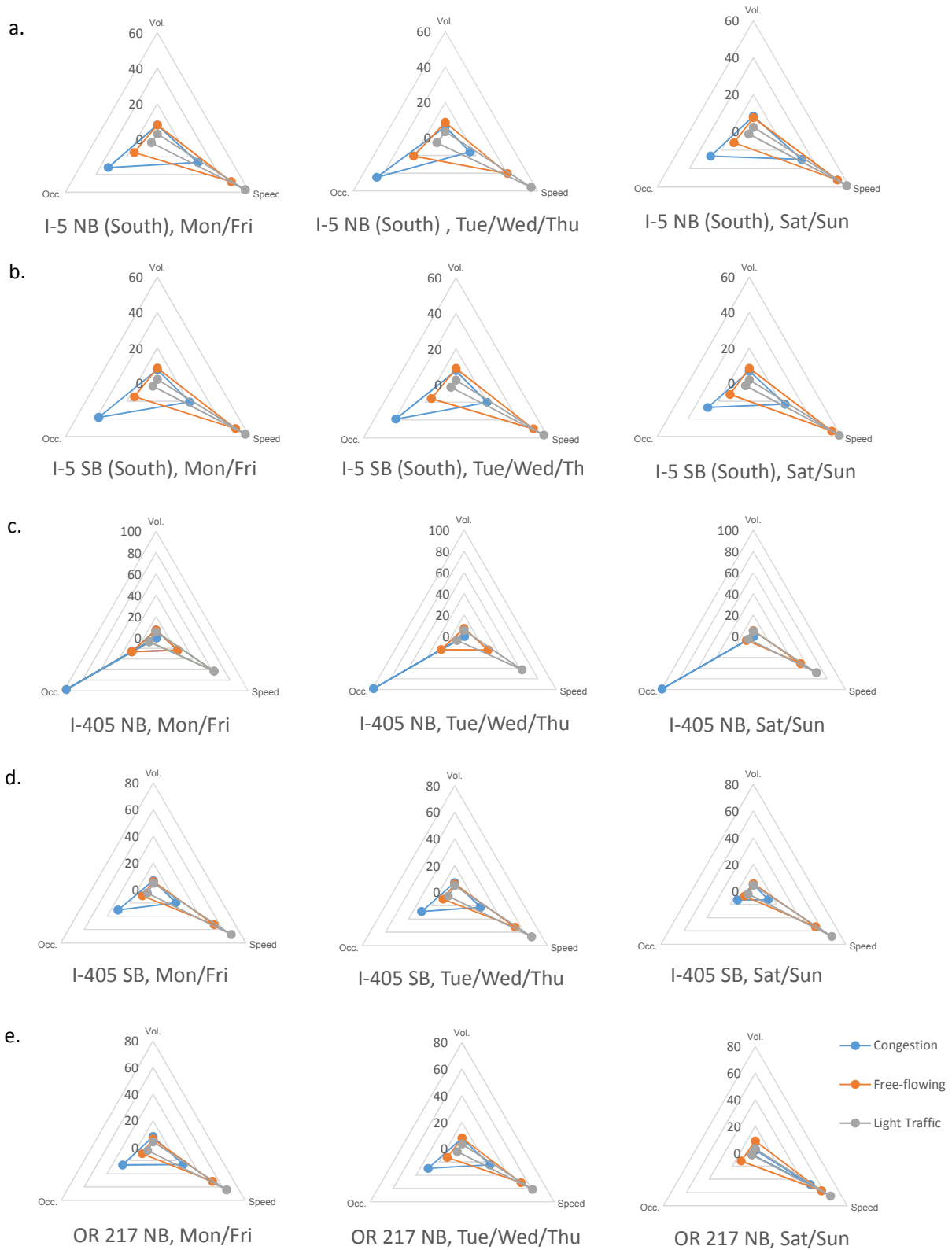

Fig. 4. Traffic Regimes by Day. Each line represents a different freeway segment, with three plots for the different days of the week: Monday/Friday, Tuesday/Wednesday/Thursday, and Saturday/Sunday. As before, each spider plot shows the k=3 clusters for an individual freeway segment, for a subset of "similar traffic days". In (c), the anomalous pattern for highway segment I-405 NB, with occupancy of 100, can clearly be seen. Note that in each case the three series are ordered from lowest to highest speed, but are not directly related from one figure to the next. (a) I-5 Northbound (South of downtown) (b) I-5 Southbound (South of downtown) (c) I-405 Northbound (d) I-405 Southbound (e) OR 217 Northbound

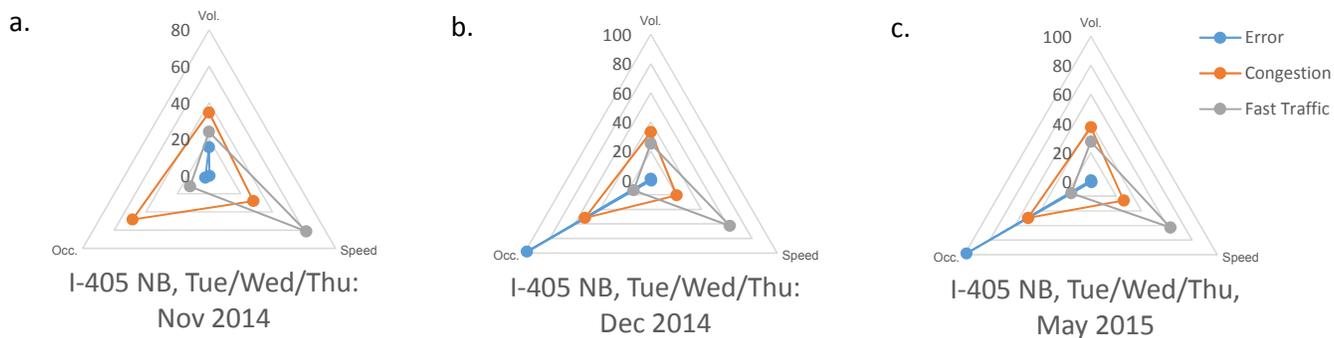

Fig. 5. Anomalous traffic regime. Plots for November (a) and December 2014 (b) show the change in the clusters as the new sensors are installed. Note that the November clusters also do not resemble the other plots, reflecting a different anomaly.

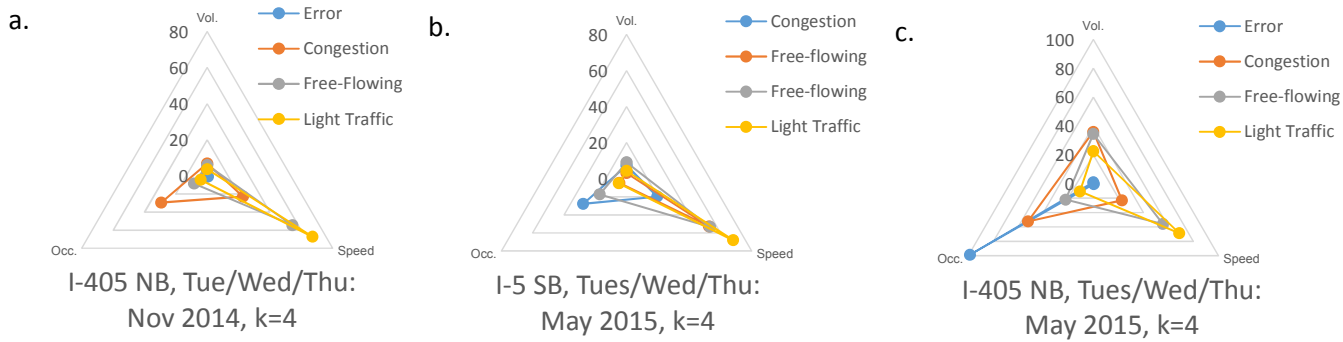

Fig. 6. k=4 plots. When reprocessed with k=4, the November (a) and May 2015 plots (c) show the same three clusters as can be seen in the plots for the other freeway segments, in addition to the anomalous cluster. Plot (b) shows a non-anomalous plot, reprocessed with k=4; here, the free-flowing cluster is broken into two smaller clusters.

For outlier detection, we calculated statistics and plotted the assigned distances across the different traffic regimes, and across some different clusterings, to ensure that the results were not significantly skewed. We also researched a randomly selected set of identified outliers and non-outliers, to ensure that their classification seemed plausible. Lastly, we calculated travel times for a section of freeway for which ground-truth was available. The ground-truth consists of actual travel times of vehicles traveling between two locations on two freeway segments, and including the month of May 2015. This data is aggregated to the one-minute level. We compared the results of travel time estimates calculated with data cleaned using the rule-based data-cleaning approach to those calculated after removing outliers identified by our anomaly-detection method.

## V. RESULTS

In this section we report results of our classification and outlier detection experiments, and of using the resulting cleaned data to calculate travel times.

### A. Classification Results

Figure 2 shows plots of the average squared error for values of *K* from 1 to 10 for the selected freeway segments. Each plot shows separate lines for all days of the week, and for each of the day-of-week groupings. The details of each chart vary. For example, the Saturday/Sunday curve is more widely separated from the other curves for I-5 North of downtown than for other freeway segments. However, in all cases, the knee of the curve falls around 3 or 4. These results are in line with the domain expert's description of 3 primary traffic regimes.

Figure 3 shows a spider plot of the cluster centers for one selected freeway segment (I-5 NorthBound (NB), South of downtown Portland), while Figure 4 shows the plots for the other segments. Note in Figure 4 that the plot for I-405 NB stands out visually as having a different pattern of clusters from the rest. We investigated the underlying data and discovered a single large cluster of bad data (51,629 entries, out of 1.345M, or 3.8% of the total), coming primarily (95% of observations in the cluster) from one detector. This particular pattern of bad data was a new pattern not currently being checked for. We have not yet established if it is a result of failing detectors, misconfiguration, or faulty ingestion procedures. Now that this pattern has been identified and recognized, it could be established as a static rule, and once we have established the likely cause and correct response, further cases can be identified and corrected.

To see how this cluster pattern developed, we went back several months to the month prior to when the sensors at detectors were replaced (see Figure 5, Nov 2014) and the month of their replacement, part-way through the month (Figure 5, Dec 2014). The December plot clearly shows the anomalous cluster, while the November plot shows a different anomalous cluster. Investigation showed a different data issue in one cluster in November. We also show in Figure 6(c) the chart for May 2015, reprocessed with 4 clusters. The three non-anomalous clusters in Figure 6(c) are similar to those in plots for other freeway segments.

While there are clear patterns visible to the domain expert, as the number of traffic segments (and spider charts) increases it may be useful to have a metric that can capture when the pattern of clusters changes significantly. Much cluster-comparison research focuses on the task of comparing different clusterings of the same dataset (e.g., [25], [26]). However, we wish to see whether the cluster centers are changing over time, ignoring differences in the details of the underlying data. In essence, we wish to have a metric that computes the similarity between two sets of cluster centers (for different datasets). We have started experimenting with a distance-based metric that sums, for each cluster center in one set, the distance to the closest cluster center in the second set. Initial tests using a simple Euclidean calculation are showing promising results; when the clusters are very similar, as in most of the spider plots in Figure 4, the distance is small. Where there are changes such as those between November and December for I-405 NB, the distance is more than 20 times larger. This measure may be a useful addition to the traffic manager's toolkit.

Figure 7 shows the data for one day, for one detector, along with a chart showing the traffic regime that each observation was assigned to. The shift from light through free-flowing traffic to peak hour congestion and back can clearly be seen. The oscillation of traffic speeds, volumes and occupancy from one observation to the next visible in the top three graphs is reflected in the oscillation between traffic regimes. Since domain experts (and drivers) tend to operate in terms of time periods rather longer than a single observation, we wish to experiment with the effect of assigning traffic regimes across a temporally defined group of observations. However, too long an average could easily hide the effects of traffic incidents or transitional periods, so caution and extensive testing against real situations is warranted. Figure 7(e) shows the effect of applying a "rolling average" over 5 observations.

### B. Outlier Detection Results

Figure 8 plots the Mahalanobis distances for each observation, for the same detector and day shown in Figure 7. The set of observations with distance $d \geq 4$, plotted in red, are clearly very different from the majority of the other observations, as are many of the observations colored in orange. Reviewing the details of these observations confirmed that they were anomalies. The frequency and timing of observations with distances between 2 and 3, here colored in yellow, caused us to speculate that many of these (normally unusual) combinations may occur during transitions from one traffic regime to another. For our subsequent tests, we chose to classify observations with $d > 2.5$ as outliers.

We found that the percentage of outliers identified in different distance categories varied significantly across the freeway segments. For outliers with a distance $d > 4$, I-205 NB had the lowest percentage (0.82%). This freeway is also the one with the large anomalous cluster. Two other freeway segments (I5 South, NB and SB) had the highest percentage (16.4% each). We speculate that these high percentages may represent anomalous clusters that are themselves large, but not large enough to swamp the "normal" clusters.

We see an opportunity to further explore additional clusterings, to see whether these clusters can be identified.

### C. Travel-Time Comparisons

To test whether removing the outliers made a discernable difference, we used the cleaned data in an existing application – calculating travel times.

First, we used the existing travel-time calculations and applied them to both traffic data cleaned using the rule-based method and data cleaned using our method. We further limited the calculations to weekdays only, and only the afternoon peak, at the request of the domain expert. We performed this comparison for a number of the freeway segments shown here. Our cleaning method produced different travel times from the rule-based method approximately 20% of the time. We explored a subset of the differences in detail. Some of them could be traced to clearly erroneous observations being excluded by our method. However, in other cases, the methods produced results that were different but both plausible, or, our method excluded observations that looked plausible.

As a second step, we calculated travel times for a segment of freeway and time period for which a "ground truth" data set was available. Again, we used only weekdays and the afternoon peak; we only included times for which we had travel times from both methods and from ground truth (since each dataset had missing entries). Overall, the methods (rounded to the nearest minute) agreed with each other 79% of the time; on average each method agreed with ground truth in 13% of the cases, but with each other in 12% of the cases. Table 1 summarizes our results.

| | NB/MF | NB/TWT | SB/MF | SB/TWT | Avg. |
|---|---|---|---|---|---|
| Both methods agree with GT | 15% | 15% | 8% | 9% | 12% |
| Methods agree, but differ from GT | 68% | 66% | 63% | 73% | 67% |
| New method only agrees with GT | 1% | 2% | 2% | 1% | 1% |
| Rule-based method only agrees with GT | 1% | 2% | 0% | 1% | 1% |
| Methods disagree with each other and with GT | 15% | 15% | 28% | 16% | 18% |
| Number of observations | 1,203 | 1,932 | 1,101 | 1,942 | |

Table 1. Agreement of afternoon peak hour travel time calculations with ground truth (GT) for freeway segment I-405.

We further investigated the 16% of cases where neither method agreed with the ground truth (a freeway segment where the average travel time is approx. 11 minutes; during the afternoon peak, the average is 12.5 minutes). In around 60% of these cases, the new method was on average 1.5 minutes closer to the ground truth, while in 40% the rule-based method was closer by around the same amount. Interestingly, the majority of the cases where the new method was closer were on the freeway segment with the anomalous cluster.

We believe that with refinement these results can be further improved, and that they warrant additional experimentation.

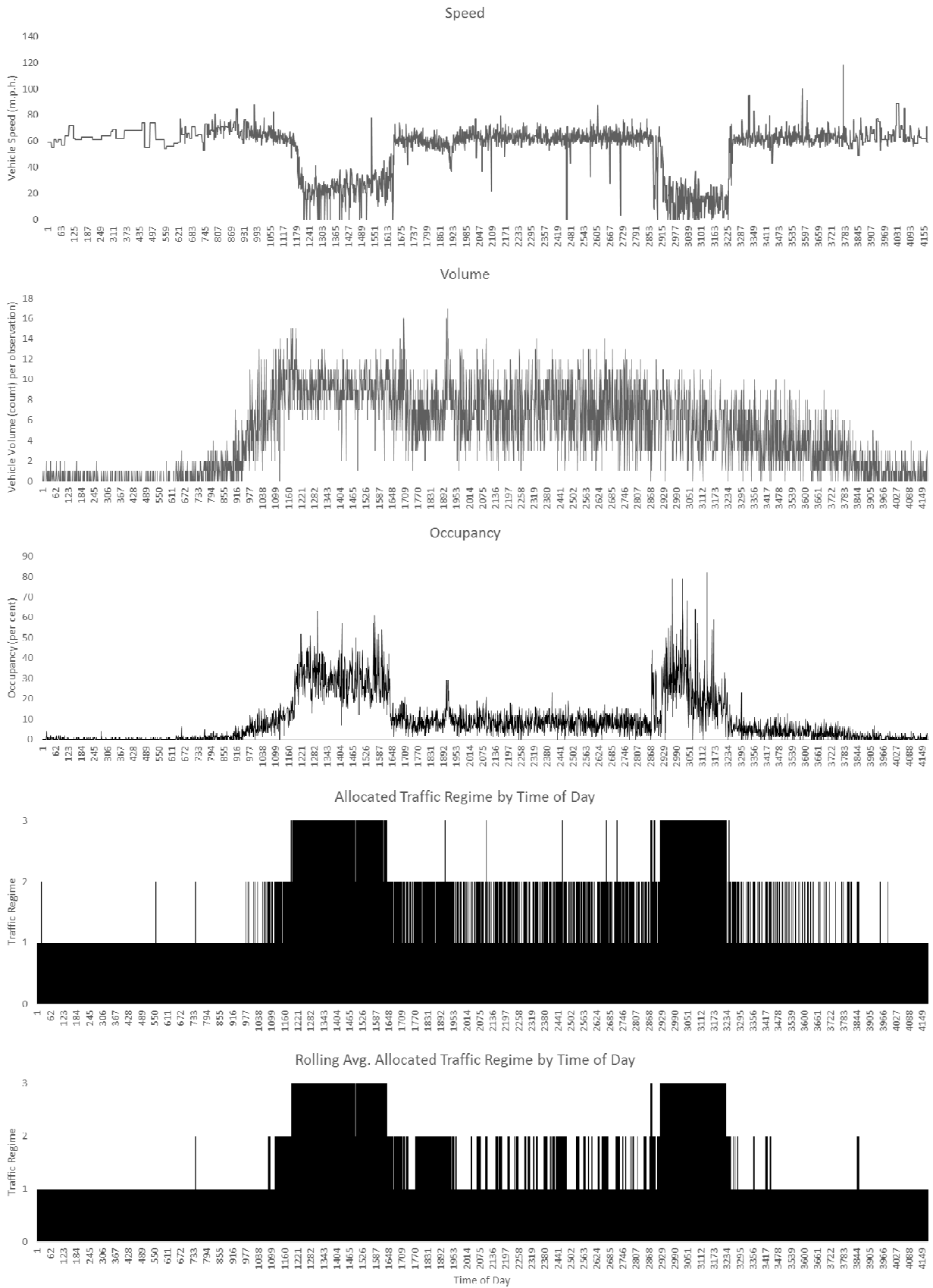

Fig. 7. Traffic Regimes. The top three charts show the source data (speed, volume and occupancy) for one detector on one day (a Wednesday), by time of day. The bottom two charts show the allocation of each observation to the "closest" cluster, representing a traffic regime (1 = light traffic; 2 = free-flowing traffic; 3 = congested). The bottom chart shows how the regime allocation changes by taking a five-observation average (approx. 1.5 minutes).

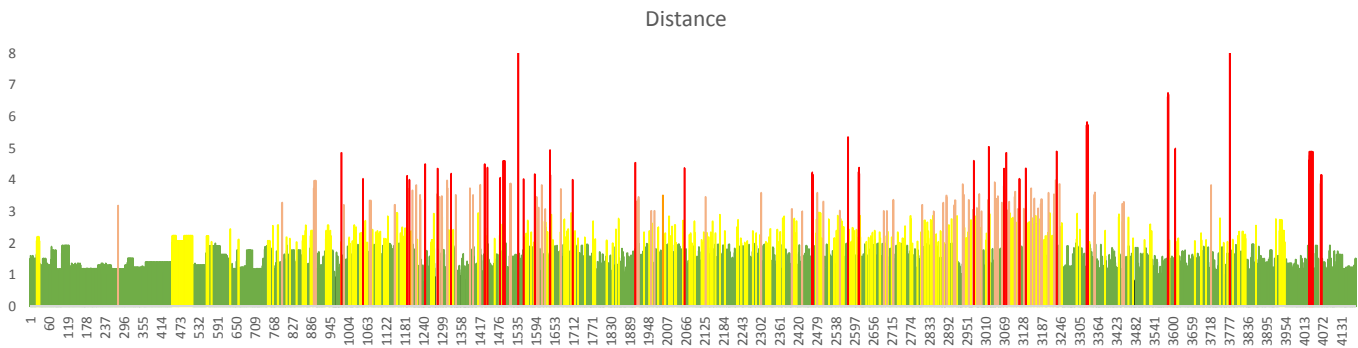

Fig. 8. Plot of Mahalanobis distance from closest centroid for one detector for one day, by time of day. Red ($d$=>4) and orange points (4>$d$=>3) are considered anomalous; yellow (3>$d$=>2) are suspect, and green ($d$<2) are considered "good".

## VI. Conclusion and Future Research

We are greatly encouraged by the results of our initial explorations. We believe that a series of plots such as those in Figure 4, provided, for example, on a regular basis, could allow traffic analysts to quickly identify potentially bad sensors or developing clusters. Reviewing the data assigned to that cluster gives the analyst a useful starting point for researching the causes. The analyst can then correct the underlying problems identified or exclude the data from further analysis, as appropriate. We have discussed including progressive calculation of clusters as a standing query in a streaming engine, and flagging anomalies on input.

There are many avenues for future research. For the clusters analysis, we would like to perform more robust cross-validation of the cluster centers identified, and to further develop a metric that can identify when the cluster composition for a freeway segment changes in a significant way.

To detect a wider range of unlikely observations, we would like to experiment with including the previous and next data points, temporally, in the anomaly detection. We want to more reliably identify cases where a detector is stuck and continues to report the same value for some period of time; in this situation, each individual report is itself a likely value, but the sequence is unlikely. Similarly, values that are too far different from the prior and next values are more likely to be the result of a misread.

The next phase is to use logistic regression to improve travel-time estimation. Factors such as weather and school holidays are widely believed to affect travel times, and including this data in the logistic regression calculations will allow us to explore their effects.

The success of this initial exploration, and understanding the application of these techniques to the current data archive, opens the door to further applications of these techniques within the traffic domain, and into other applications that support sustainable development.

We believe that applying machine-learning techniques to sensor observations adds to our ability to interpret and model the socioeconomic-environmental systems observed by the sensors. By reflecting the current mental models of domain experts (in this case, by statistically identifying traffic regimes), and then further applying that knowledge to improving data quality (here, by identifying anomalous traffic, detectors, or individual outlier observations), we allow domain experts to more confidently identify, explore and predict changes to the systems themselves.

## Acknowledgment

We thank the Oregon Department of Transportation for their support.